\documentclass[%
reprint,
superscriptaddress,
 amsmath,amssymb,
 aps,
pra,
]{revtex4-2}

\usepackage{graphicx}% Include figure files
\usepackage{dcolumn}% Align table columns on decimal point
\usepackage{bm}% bold math
\usepackage{hyperref}% add hypertext capabilities
\usepackage{tabularx,booktabs}% For tables
\usepackage{soul}
\hypersetup{
    colorlinks=true,
    linkcolor=black,
    citecolor=blue,
    urlcolor=black
    }

\usepackage{color}
\newcommand{\revisedtext}[1]{{\color{black}#1}}

\usepackage{algorithm} 
\usepackage{algpseudocode} 

\usepackage{amsmath}

\begin{document}

\preprint{APS/123-QED}

\title{Recovery of Quantum Correlations using Machine Learning}

\author{Edward W. Steele}
\affiliation{Department of Electrical Engineering, University of Tennessee at Chattanooga, Chattanooga, TN 37403, USA}
\affiliation{UTC Research Institute, University of Tennessee at Chattanooga, Chattanooga, TN 37403, USA}%

\author{Donald R. Reising}
\email{donald-reising@utc.edu}
\affiliation{Department of Electrical Engineering, University of Tennessee at Chattanooga, Chattanooga, TN 37403, USA}
\affiliation{UTC Research Institute, University of Tennessee at Chattanooga, Chattanooga, TN 37403, USA}%
\affiliation{UTC Quantum Center, University of Tennessee at Chattanooga, Chattanooga, TN 37403, USA}

\author{Tian Li}
\email{tian-li@utc.edu}
%\homepage{T.L. and Z.J. contributed equally to this work.}
\affiliation{UTC Research Institute, University of Tennessee at Chattanooga, Chattanooga, TN 37403, USA}
\affiliation{UTC Quantum Center, University of Tennessee at Chattanooga, Chattanooga, TN 37403, USA}
\affiliation{Department of Chemistry and Physics, University of Tennessee at Chattanooga, Chattanooga, TN 37403, USA}


\begin{abstract}
Quantum sources with strong correlations are essential but delicate resources in quantum information science and engineering. Decoherence and loss are the primary factors that degrade nonclassical quantum correlations, with scattering playing a role in both processes. In this work, we present a method that leverages Long Short-Term Memory (LSTM), a machine learning technique known for its effectiveness in time-series prediction, to mitigate the detrimental impact of scattering in quantum systems. Our setup involves generating two-mode squeezed light via four-wave mixing in warm rubidium vapor, with one mode subjected to a scatterer to disrupt quantum correlations. Mutual information and intensity-difference squeezing between the two modes are used as metrics for quantum correlations. We demonstrate a 74.7~\% recovery of mutual information and 87.7~\% recovery of two-mode squeezing, despite significant photon loss that would otherwise eliminate quantum correlations. This approach marks a significant step toward recovering quantum correlations from random disruptions without the need for hardware modifications, paving the way for practical applications of quantum protocols.
\end{abstract}

\maketitle

%\tableofcontents

\section{\label{sec:level1}Introduction}

Quantum information science and engineering (QISE) leverages the core principles of quantum mechanics to handle and process information in ways that outperform classical techniques, paving the way for significant advancements in quantum computing, communication, and sensing~\cite{nielsen2010quantum}. Central to these breakthroughs are quantum correlations, which provide the non-classical resources essential for QISE. However, these correlations are intrinsically fragile and highly sensitive to environmental disturbances. Disruptive processes, namely decoherence and loss, can severely degrade these correlations, resulting in the loss of the unique quantum behaviors that distinguish QISE systems from their classical counterparts~\cite{joos2013decoherence}. As a result, addressing these processes' adverse effects is paramount for practically realizing QISE protocols. %the adverse effects of these processes is of paramount importance for the practical realization of QISE protocols. 
Strategies to protect quantum correlations from disruptive environments are crucial for deploying robust quantum systems in real-world applications~\cite{bachor2019guide}. Thus, successfully mitigating decoherence and loss will enhance the stability of quantum systems and expedite the deployment of QISE protocols in real-world settings.

Among the most extensively studied quantum systems, squeezed states of light have consistently demonstrated their utility as reliable quantum resources for numerous QISE protocols~\cite{li2022quantum,casacio2021quantum,de2020quantum,lawrie2019quantum,michael2019squeezing,clark2016observation,aasi2013enhanced,taylor2013biological}, especially in the continuous variable (CV) regime~\cite{braunstein2005quantum}. Complementary to our recent work~\cite{PRXQuantum.5.030351}, where we demonstrated the mitigation of scattering effects using a simple hardware solution—an integrating sphere (IS)—we now turn our attention to a software-based approach to mitigate the adverse effects caused by scattering. This is particularly crucial for QISE as scattering contributes to both decoherence and loss, the two key challenges in maintaining quantum coherence and correlations~\cite{nielsen2010quantum}. This study uses a two-mode squeezed state of light produced through the four-wave mixing (FWM) process in warm rubidium vapor as the quantum source to explore how machine learning (ML) algorithms can mitigate the disruptive effects on quantum correlations. While effective quantum correlation restoration schemes, such as active shaping of twin photons' temporal wave packets to revive entanglement by restoring quantum interference~\cite{Wu2019}, show promise, they face significant challenges in real-world applications, where absorption and random scattering lead to severe losses.

%Additionally, although ML framework based on Bayesian inference has been employed to suppress excess noise and enhance the secret key rate of continuous-variable quantum key distribution (CV-QKD)~\cite{chin2021machine}, no demonstrations have yet shown ML's potential to recover quantum correlations from disruptive processes. 

\textcolor{black}{Recent advances have demonstrated the utility of ML techniques in quantum information science, such as continuous-variable quantum key distribution (CV-QKD)~\cite{chin2021machine}, quantum error correction~\cite{nautrup2019optimizing}, quantum control~\cite{niu2019universal}, and particularly in quantum state tomography (QST)~\cite{innan2024quantum}, which aims to reconstruct the \textit{complete quantum state} of a system, represented by its density matrix, based on experimental measurements. This process is crucial for verifying quantum system behavior, characterizing quantum devices, analyzing quantum phenomena such as entanglement and coherence, and benchmarking experimental results against theoretical predictions~\cite{cramer2010efficient,lvovsky2009continuous}. Machine-learning-based QST (ML-QST) has emerged as a transformative approach to quantum state reconstruction, addressing key challenges of scalability and noise in traditional QST~\cite{torlai2018neural}. Leveraging advanced ML techniques such as convolutional neural networks~\cite{schmale2022efficient,lohani2020machine}, adaptive neural networks~\cite{quek2021adaptive}, deep learning~\cite{koutny2022neural}, and most recently quantum machine learning~\cite{innan2024quantum}, ML-QST enables efficient processing of large datasets, reduces measurement requirements, and enhances noise resilience~\cite{palmieri2024enhancing}. It is versatile, applicable across diverse quantum platforms, and effective at handling high-dimensional systems~\cite{quek2021adaptive}. \textcolor{black}{Unlike ML-QST, which reconstructs a \textit{complete static state} of a quantum system, our approach focuses on \textit{recovering dynamic quantum correlations within a quantum system, rather than reconstructing the quantum state itself}, based on partial or indirect information. Specifically tailored for \textit{evolving} scattering scenarios, our method employs \textit{time-series predictions} to restore quantum correlations in a quantum system, offering a distinct and complementary perspective, which also differentiates it from static beamsplitter-modeled analytical loss formulas.}}

In this work, we introduce a scatterer—a ground glass diffuser—to one of the two modes of the squeezed state to simulate a realistic disruptive environment. In contrast to the hardware-based approach detailed in Ref.~\cite{PRXQuantum.5.030351}, this work achieves mitigation using Long Short-Term Memory (LSTM), a recurrent neural network (RNN) architecture in ML~\cite{hochreiter1997long}. Using the time sequence difference between the two quantum-correlated modes before the scatterer was introduced and the time sequence of the undisrupted mode, the LSTM model is employed to reconstruct the disrupted mode, effectively restoring the quantum correlations.

\begin{figure*}
\centering
\includegraphics[width=\linewidth]{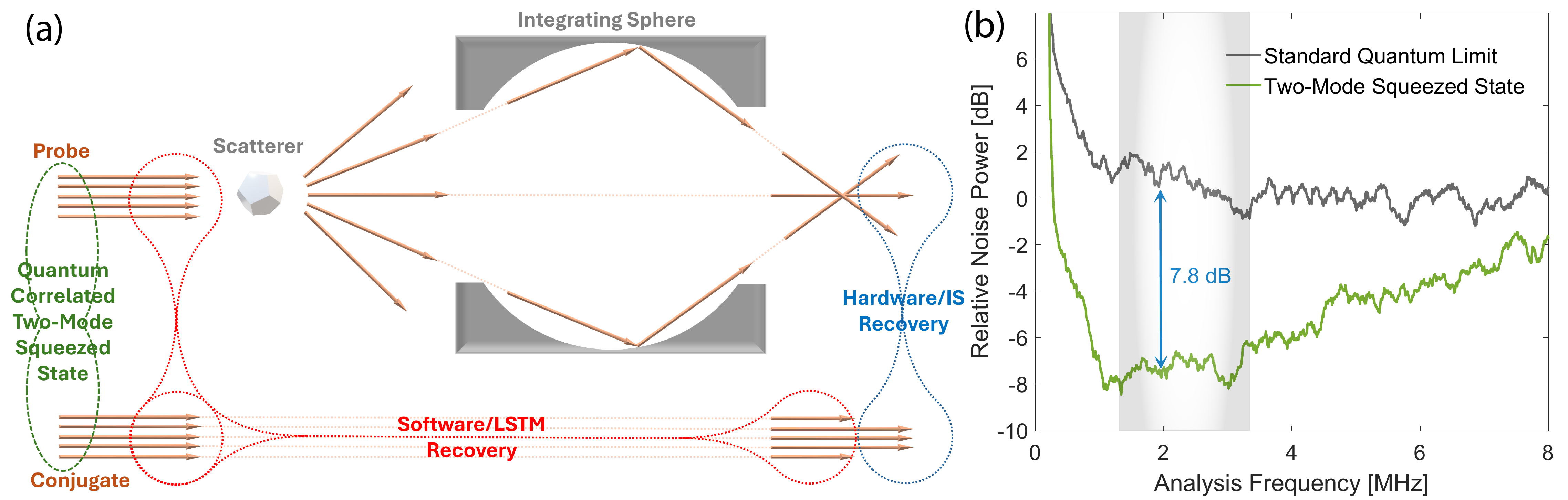}
\caption{(a) Schematic that depict the hardware-based scattering mitigation approach in Ref.~\cite{PRXQuantum.5.030351} and the software/LSTM-based approach in this work. The scatterer is introduced \textit{only to the probe beam}. (b) A typical spectrum of two-mode intensity-difference squeezing is obtained by post-processing the intensity fluctuations of the probe and conjugate beams using Matlab. The green and gray curves represent the spectra of the squeezed and coherent (i.e., shot-noise limited, or SQL) beams, respectively. The shaded region between 1.5~MHz and 3.5~MHz indicates our focus area for data analysis.
\label{fig1}}
\end{figure*}

Following Ref.~\cite{PRXQuantum.5.030351}, we use mutual information (MI)~\cite{serafini2003symplectic} as a metric for assessing quantum correlations. Both MI and its quantum counterpart, quantum mutual information (QMI), are proven effective for quantifying these correlations~\cite{clark2014quantum,vogl2014advanced,weedbrook2012gaussian}. MI measures the shared information between two correlated entities, providing valuable insights into their overall correlations, including quantum correlations~\cite{nielsen2010quantum}. QMI has been utilized in the CV regime for bipartite Gaussian states to represent information coherently within quantum systems, enabling some important quantum information processing implementations~\cite{ameri2015mutual,dixon2012quantum,wolf2008area,belavkin2002entanglement}. However, it is well known that to calculate QMI for bipartite Gaussian states, one must first access the covariance matrix. Since our work does not involve measuring field quadrature, we opted not to use QMI to measure quantum correlations. Instead, we compute MI using Shannon entropy~\cite{serafini2003symplectic} for our quantum system, which consists of quantum-correlated bright two modes generated through the FWM process. Specifically, \textit{the MI is derived from the joint probability of the intensity fluctuations of the two modes, encapsulating their quantum squeezed nature}. By applying our LSTM-based scattering mitigation scheme, we successfully recover 74.7~\% of MI, a substantial improvement over the 47.4~\% recovery reported in Ref.~\cite{PRXQuantum.5.030351}. Additionally, we recover 87.7~\% of the two-mode squeezing despite significant photon loss that would otherwise destroy quantum correlations. The simple and straightforward nature of our ML-aided approach renders it particularly well-suited for QISE protocols in which disruptive processes are inevitable.

\begin{figure*}
\centering
\includegraphics[width=\linewidth]{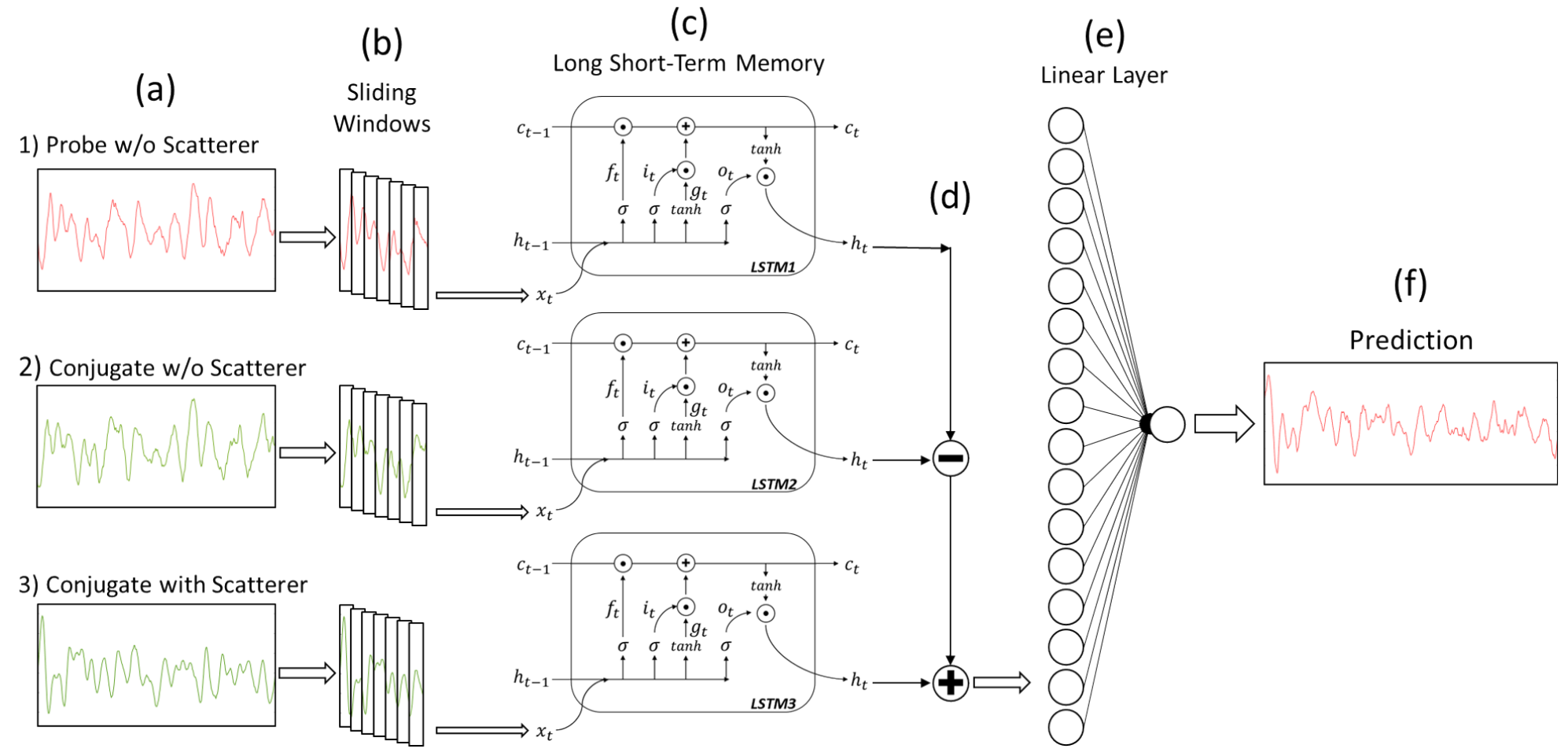}
\caption{LSTM flowgraph. (a) the input quantum data: 1) the probe time sequence before the scatterer is introduced, 2) the conjugate time sequence before the scatterer is introduced, and 3) the conjugate time sequence after the scatterer is introduced, labeled as 'Probe w/o Scatter,' 'Conjugate w/o Scatterer,' and 'Conjugate with Scatterer,' respectively. (b) the data preparation: sliding windows, (c) the LSTM layer with three LSTM blocks, (d) the subtraction and addition operation, (e) the fully connected linear layer, and (f) the output prediction: the reconstructed disrupted probe time sequence.
\label{fig5}}
\end{figure*}

\section{Experiment}

The two-mode squeezed light utilized in this study is generated through the FWM process in a warm atomic vapor cell of $^{85}$Rb~\cite{li2022quantum,dowran2018quantum,anderson2017phase,pooser2015ultrasensitive,clark2014quantum,hudelist2014quantum}. This generation scheme has proven effective for producing quantum correlations~\cite{li2021experimental,li2017improved,clark2014quantum,vogl2014advanced,vogl2013experimental}. A comprehensive description of the experimental setup can be found in Ref.~\cite{PRXQuantum.5.030351}.

Figure~\ref{fig1}(a) illustrates a schematic that compares the hardware-based approach detailed in Ref.~\cite{PRXQuantum.5.030351} with the ML-aided method employed in this work. Since photons are always produced in pairs, hence the number variation of generated photons, i.e., the intensity fluctuations of the two beams—designated as the `probe' and `conjugate' beams in Fig.~\ref{fig1}(a), collectively referred to as the `twin beams'—exhibit strong quantum correlations, evidenced by significant two-mode intensity-difference squeezing. A typical squeezing spectrum, derived from post-processing the twin beams' intensity fluctuations, is shown in Fig.~\ref{fig1}(b). The standard quantum limit (SQL), represented by the grey curve, was measured using the methodology described in Ref.~\cite{PRXQuantum.5.030351}. The green curve in Fig.~\ref{fig1}(b) indicates the noise power for the intensity difference between the twin beams, demonstrating a 7.8~dB reduction (squeezing) below the SQL.

In this study, we focus on measuring squeezing in the intensity difference of a bipartite entangled state, rather than in field quadratures. This entangled state is created through the FWM process in $^{85}$Rb, where two photons from a pump beam are transformed into twin photons emitted into spatially separate probe and conjugate modes. Known as ``bright two-mode squeezing," this process yields optical power for both the probe and conjugate beams ranging from nano-watt to mili-watt~\cite{li2020squeezed}. While photon generation in each mode is random and exhibits thermal-like characteristics, strong quantum correlations exist between the intensity fluctuations of the probe and conjugate modes, as the entangled photons are generated in pairs through the FWM process. These strong correlations lead to the observed squeezing in the intensity difference between the probe and conjugate beams, as depicted in Fig.~\ref{fig1}(b).

As illustrated in Fig.~\ref{fig1}(a), the conjugate beam travels without disruption. In contrast, the probe beam passes through a scatterer, i.e., a half-inch diameter ground glass diffuser, with its polished surface oriented toward the beam. The reverse side of the scatterer has a 120-grit medium surface, which generates near-spherical scattering patterns. In the hardware-based scattering mitigation approach described in Ref.~\cite{PRXQuantum.5.030351}, the scatterer is positioned at the entrance aperture (also half-inch diameter) of a 5-cm-diameter integrating sphere (IS), allowing nearly all scattered photons to be recollected. In the ML-aided scattering mitigation approach explored here, the LSTM architecture uses the \textit{difference} between the time sequences of the twin beams before the scatterer was introduced and the time sequence of the undisrupted conjugate beam as input. This is illustrated by the red dashed lines in Fig.~\ref{fig1}(a).

We used an oscilloscope to record photo-currents from the photo-detectors that capture the photon streams in the twin beams. This enables post-processing the quantum correlations in the fluctuations of their photon numbers, i.e., the fluctuations in their intensities. The recorded photo-current, i.e., oscilloscope time sequence, from each detector, consists of 4 million data points, sampled at a rate of 2.0~giga-samples per second, covering a total acquisition time of 2~ms. In the following sections, we employ the MI metric to characterize overall correlations, including quantum correlations, between the twin beams and introduce a Long Short-Term Memory (LSTM) architecture to recover these correlations by reconstructing the time sequence of the disrupted probe beam.

\section{LSTM Architecture%
\label{sec:lstm_arch}}

We use the quantum-correlated probe and conjugate time sequences recorded by the oscilloscope as the input data for our ML algorithm. Specifically, the three inputs to our LSTM architecture, shown in Fig.~\ref{fig5}(a), are 1) the probe time sequence before the scatterer is introduced, 2) the conjugate time sequence before the scatterer is introduced, and 3) the conjugate time sequence after the scatterer is introduced, labeled as 'Probe w/o Scatter,' 'Conjugate w/o Scatterer,' and 'Conjugate with Scatterer,' respectively in Fig.~\ref{fig5}(a). The time sequences are prepared through sliding windows of a specific window length, which slide across the time sequence in single-point iterations from the beginning to the end of the time sequence as illustrated in Fig.~\ref{fig5}(b). %This is illustrated in Fig.~\ref{fig5}(b). 
The number of sliding windows equals the total length of the time sequence minus the window length minus 1. The data contained in each sliding window is the input to the LSTM layer.

Within the LSTM layer shown in Fig.~\ref{fig5}(c), the input state at time $t$, $x_t$, is concatenated with an initial zero hidden state at $t-1$, $h_{t-1}$, of a specific hidden length. The concatenated input and hidden states go through four separate gates, i.e., neural networks, which are the forget gate, $f_t$, the input gate, $i_t$, the cell gate, $g_t$, and the output gate, $o_t$. Gate operations consist of weight matrices, bias vectors, and activation functions. The forget, input, cell, and output gates are given in Eqs.~\eqref{eq2},~\eqref{eq3},~\eqref{eq4},~\eqref{eq5}, respectively:
\begin{eqnarray}
    f_t = \sigma(W_f [h_{t-1}, x_t]+b_f),
    \label{eq2}
\\
    i_t = \sigma(W_i [h_{t-1}, x_t]+b_i),
    \label{eq3}
\\
    g_t = \tanh(W_g [h_{t-1}, x_t]+b_g),
    \label{eq4}
\\
    o_t = \sigma(W_o [h_{t-1}, x_t]+b_o),
    \label{eq5}
\end{eqnarray}
where $\sigma$ is the sigmoid activation function, $\tanh$ is the hyperbolic tangent activation function, $W_f$, $W_i$, $W_g$ and $W_o$ are the forget, input, cell, and output weight matrices respectively, and $b_f$, $b_i$, $b_g$ and $b_o$ are the forget, input, cell, and output bias vectors respectively. The cell state at time $t$, $c_t$, and the hidden state at time $t$, $h_t$ are given in Eqs.~\eqref{eq6},~\eqref{eq7}, respectively:
\begin{eqnarray}
    c_t = f_t \odot c_{t-1}+i_t \odot g_t,
    \label{eq6}
\\
    h_t = o_t \odot \tanh(c_t),
    \label{eq7}
\end{eqnarray}
where $\odot$ is the Hadamard product, i.e., element-wise multiplication, and $c_{t-1}$ is the initial zero cell state at time $t-1$. The output of the LSTM layer is $h_t$. The $h_t$ of the `Probe w/o Scatterer' time sequence is subtracted from the $h_t$ of the `Conjugate w/o Scatterer' time sequence. This difference is added to the $h_t$ of the `Conjugate with Scatterer' time sequence. The result is the target feature hidden state, which will be used as the input to the fully connected linear layer neural network shown in Fig.~\ref{fig5}(d). The linear layer performs a linear transformation on the hidden state to reduce the size of the hidden state to a single state depicted in Fig.~\ref{fig5}(e). The output of the linear layer is the final prediction of our LSTM architecture, which is the reconstruction of the disrupted probe time sequence displayed in Fig.~\ref{fig5}(f).

The LSTM is built upon the Python library, $PyTorch$~\cite{paszke2019pytorch}. The data is split into training (64\%), validation (16\%), and testing (20\%) sets. In each set, the data is grouped in batches and randomly shuffled. The loss function is the mean squared error (MSE). The \revisedtext{stochastic} optimization method \revisedtext{to update the LSTM parameters} is ``adaptive moment estimation'' (a.k.a., $Adam$) detailed in~\cite{kingma2014adam}. The hyperparameters are given in Tab.~\ref{tab:hyper}. For every epoch, the LSTM is trained and evaluated. The training of the LSTM is conducted in the following way: 1) the LSTM is set to train mode, 2) the forward pass, 3) the loss function calculation, 4) the backward propagation, 5) the optimizing step, and 6) repeat for each batch. The time-series validation of the LSTM is conducted in the following way: 1) the LSTM is set to evaluation mode, 2) the forward pass, 3) the loss function calculation, and 4) repeat for each batch. The trained and validated LSTM is tested with the never-before-seen testing set in batches with just the forward pass for the final prediction. \revisedtext{The pseudo-code of the training and validation of the LSTM is provided in Algorithm 1. The input is the data contained in the sliding windows and the target is the next single value immediately following each window.}\\

\begin{algorithmic}[0]
    \State \textbf{\revisedtext{Algorithm 1: LSTM Training and Validation}}
    \For {$epoch=1,2,\ldots,i$}
	\For {$trainingset=1,2,\ldots,N$}
            \State \textbf{Forward Pass}
            \State $input, target\leftarrow training set$
		\State $output \leftarrow $LSTM$(input)$
            \State $Loss \leftarrow $ MSE($output$, $target$)
            \State \textbf{Backward Pass}
            \State Calculate gradients
            \State \textbf{Adam Optimization}
            \State Update LSTM parameters
	\EndFor
        \For {$validationset=1,2,\ldots,M$}
            \State \textbf{Forward Pass}
            \State $input, target\leftarrow validation set$
		\State $output \leftarrow $LSTM$(input)$
            \State $Loss \leftarrow $ MSE($output$, $target$)
        \EndFor
    \EndFor
\end{algorithmic}

\begin{table}[!h]
\centering
\caption{List of Hyperparameters.\label{tab:hyper}}
\resizebox{0.55\columnwidth}{!}{%
\begin{tabular}{cc}
\toprule
Hyperparameter & Range \\
\hline
Window Length & {[2, 100]} \\
Hidden Length & {[2, 64]} \\
Batch Size & {[100, 1000]}\\
Number of Epochs & {[10, 100]}\\
Learning Rate & {[0.1, 0.001]}\\
\bottomrule
\end{tabular}
}
\end{table}

\begin{figure}
\centering
\includegraphics[width=\columnwidth]{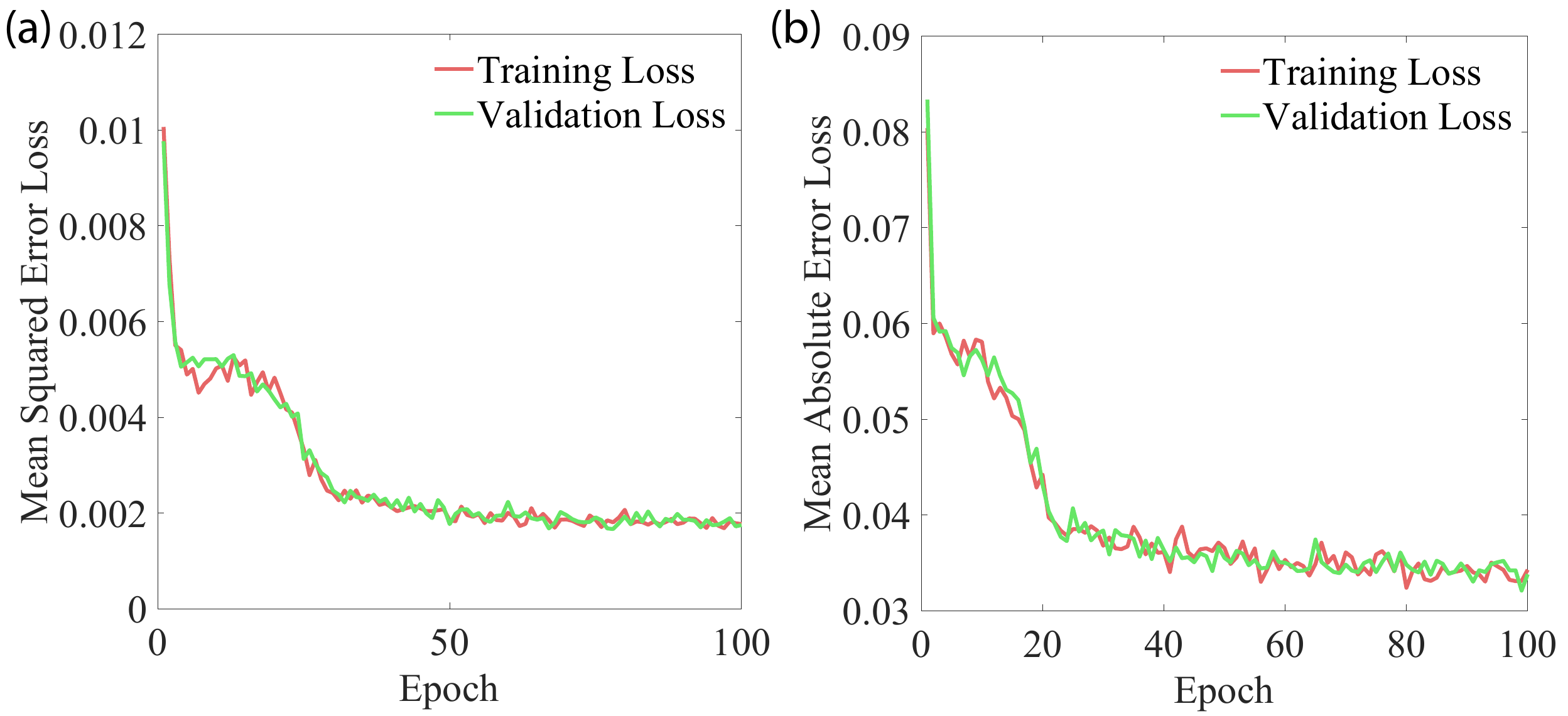}
\caption{\revisedtext{Training and validation loss curves. The convergence of the training and validation loss curves prove two things: 1) the LSTM's ability to generalize over new data, and 2) no overfitting. (a) Mean Squared Error (MSE) Loss Function. The final training loss is 0.00177. The final validation loss is 0.00175. (b) Mean Absolute Error (MAE) Loss Function. The final training loss is 0.0343. The final validation loss is 0.0338.}
\label{fig:lossfunc}}
\end{figure}

\revisedtext{To address any indication of overfitting, the loss function training and validation curves are shown in Fig.~\ref{fig:lossfunc}. Two loss functions were tested: 1) mean squared error (MSE) in Fig.~\ref{fig:lossfunc}(a), and 2) mean absolute error (MAE) in Fig.~\ref{fig:lossfunc}(b).  If the LSTM was overfitting on the training data, the validation loss should be much higher than the training loss. Since the training and validation losses converge after 100 epochs, the LSTM is able to generalize on new data and does not overfit.}

\begin{figure*}
\centering
\includegraphics[width=\linewidth]{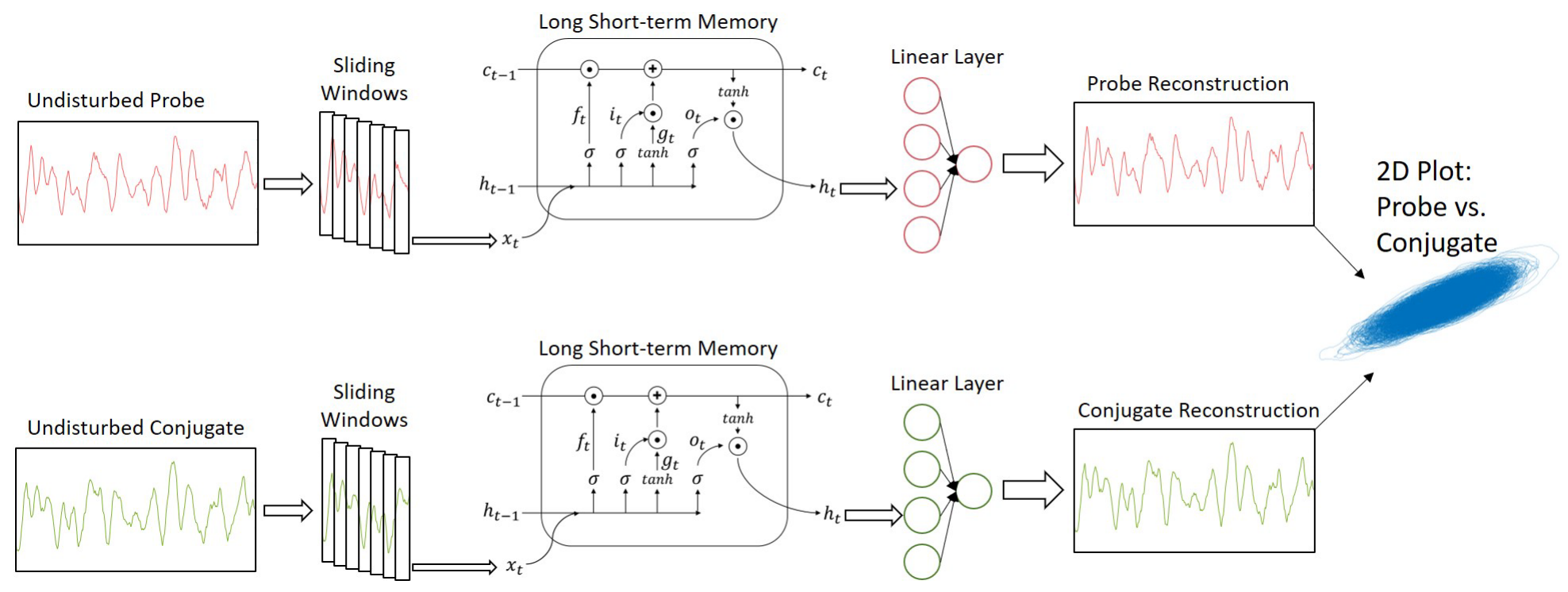}
\caption{\revisedtext{The LSTM flowgraph for individual time trace reconstruction illustrates the evaluation performance of the LSTM architecture. It is expected to effectively reconstruct the \textit{undisturbed} probe and \textit{undisturbed} conjugate time traces. Following the individual reconstructions, a 2D plot of the joint probability distribution of the reconstructed probe and conjugate time traces is presented to validate the effectiveness of the LSTM architecture.}
\label{fig7}}
\end{figure*}

\revisedtext{The complete architecture of the LSTM model to reconstruct the disrupted probe time sequence is shown in Fig.~\ref{fig5}. To prove the effectiveness and confidence of the reconstruction, a simpler, single-layer, LSTM architecture was tested on individual input time sequences: the probe and conjugate before scattering, as shown in Fig.~\ref{fig7}}. Please note that the results obtained in this work are contingent upon the conjugate time sequence after the scatterer was introduced to the probe beam path. A prediction of the probe time sequence prior to scattering would not exhibit a high correlation with the conjugate sequence post-scatterer. However, the relationship between the probe and conjugate sequences before scattering can be effectively mapped to the conjugate time sequence after scattering, enabling the prediction of the probe sequence under post-scattering conditions. It is also important to emphasize that this prediction is finite and limited by the length of the input time sequence. Infinite prediction is beyond the scope of this study.

\begin{figure*}
\centering
\includegraphics[width=\linewidth]{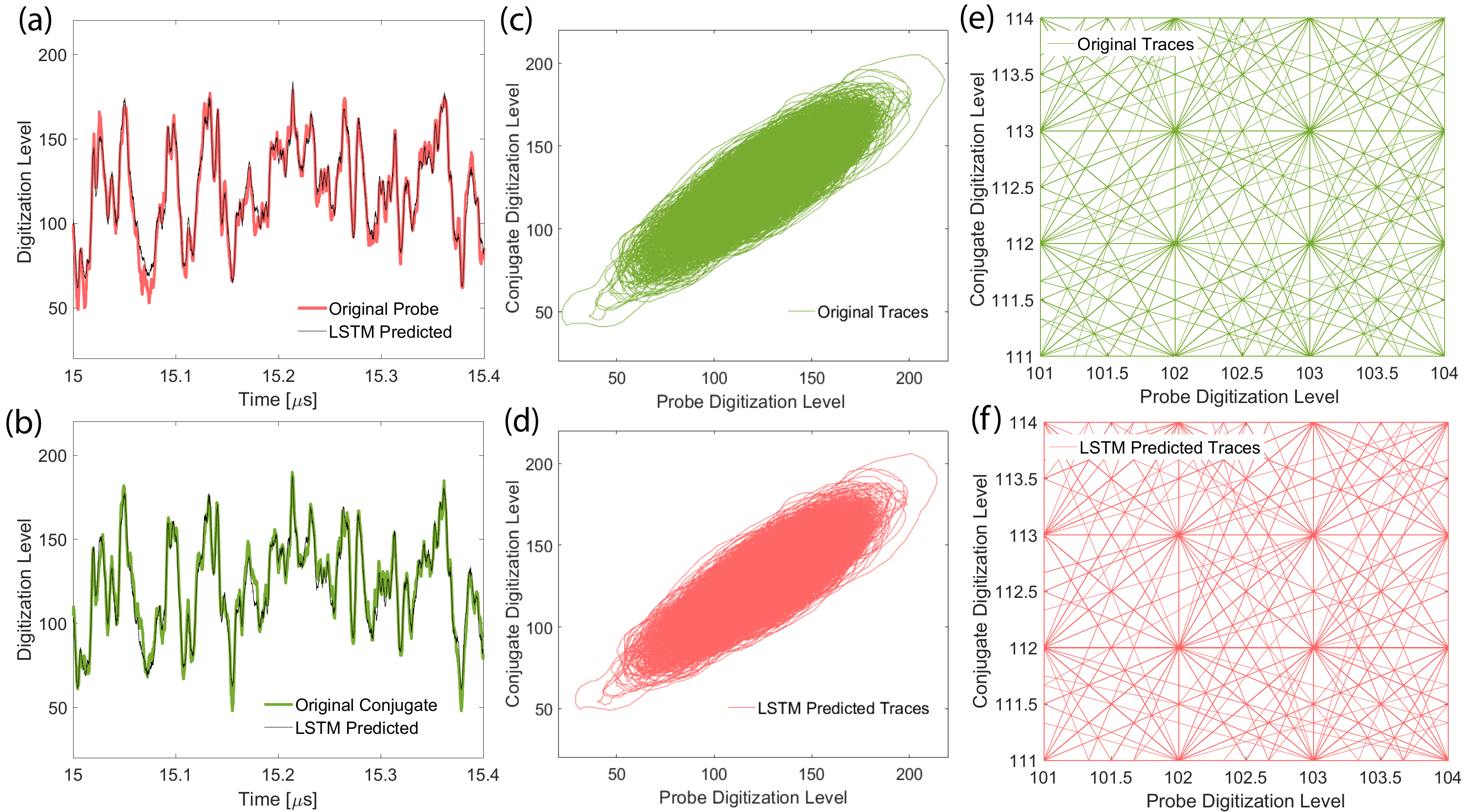}
\caption{Evaluation of the performance of our ML algorithm. Panels (a) and (b) display the time sequences of the intensity fluctuations for the undisrupted probe and conjugate beams, represented by the red and green curves, respectively. The LSTM-predicted time sequences are shown as black curves in both panels. Panels (c) and (d) display the joint probability distributions of the intensity fluctuations for the undisrupted probe and conjugate beams in (c) and the LSTM-predicted twin beams in (d). Panels (e) and (f) provide zoomed-in views of (c) and (d), respectively.
\label{fig:verification}}
\end{figure*}

\section{Results}

\subsection{Performance Evaluation of the ML Algorithm}

\revisedtext{To demonstrate the effectiveness and confidence of our LSTM architecture, we applied the algorithm to predict the \textit{undisrupted} probe and conjugate time sequences independently shown in Fig.~\ref{fig7}. This was done using the same architecture shown in Fig.~\ref{fig5}, but with only an undisrupted probe or undisrupted conjugate as the input data and a single LSTM block while skipping step (d) in Fig.~\ref{fig5}. The results are shown in Fig.~\ref{fig:verification}.} Figures~\ref{fig:verification}(a) and~\ref{fig:verification}(b) display the time sequences of the intensity fluctuations for the undisrupted probe and conjugate beams, represented by the red and green curves, respectively. The LSTM-predicted time sequences are shown as black curves in both panels. The $y$-axis in Figs.~\ref{fig:verification}(a) and~\ref{fig:verification}(b) corresponds to the 8-bit oscilloscope's the digitization level, with a maximum value of $2^8 = 256$. Figures~\ref{fig:verification}(c) and~\ref{fig:verification}(d) display the distributions of the joint probability of the intensity fluctuations for the undisrupted twins beams in Fig.~\ref{fig:verification}(c), and the LSTM-predicted twin beams in Fig.~\ref{fig:verification}(d). Figures~\ref{fig:verification}(e) and~\ref{fig:verification}(f) provide closer views of Figs.~\ref{fig:verification}(c) and~\ref{fig:verification}(d) respectively to highlight the entangled, i.e., densely connected, nature of the quantum-correlated two beams. The clear similarities between the joint probability patterns in Figs.~\ref{fig:verification}(c) and~\ref{fig:verification}(d), as well as in Figs.~\ref{fig:verification}(e) and~\ref{fig:verification}(f), manifest the satisfactory performance of our ML algorithm.

\begin{figure*}
\centering
\includegraphics[width=\linewidth]{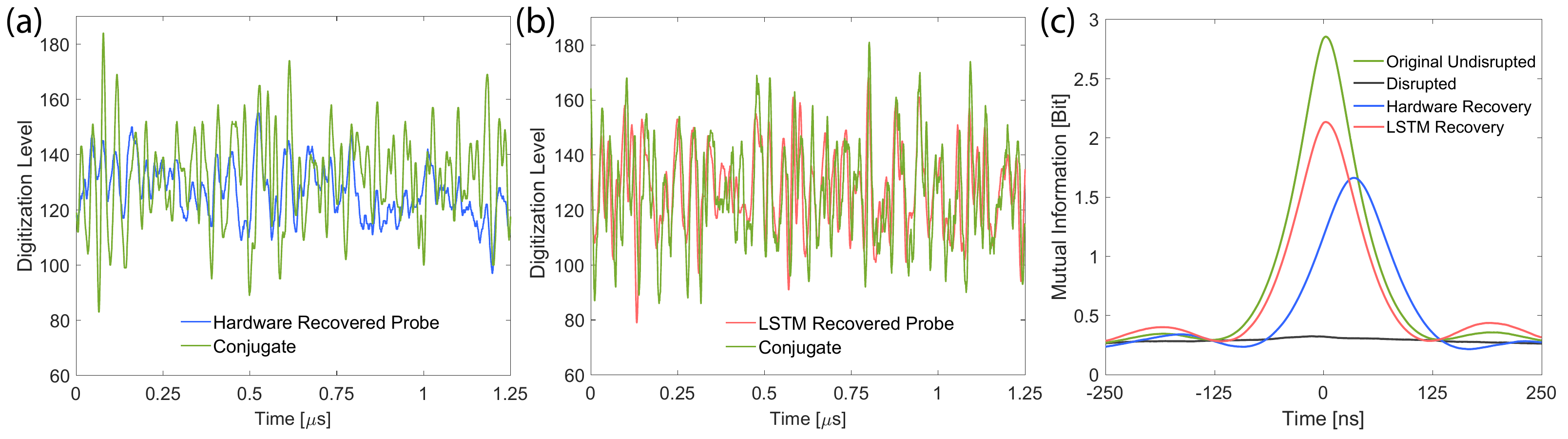}
\caption{Panel (a) illustrates the performance of the hardware-based scattering mitigation scheme using an IS to recollect scattered probe photons, as detailed in Ref.~\cite{PRXQuantum.5.030351}. Intensity fluctuations of the original conjugate beam and the recovered probe beam are shown as the green and blue curves, respectively. Panel (b) shows the performance of the ML-aided scattering mitigation scheme using an LSTM architecture to reconstruct the scattered probe time sequence, with intensity fluctuations of the recovered probe beam and the conjugate beam (with the scatterer present) depicted as the red and green curves, respectively. Panel (c) compares the two recovery schemes using MI as the information metric. The green, black, blue, and red curves represent the original undisrupted, disrupted, hardware-based recovery scheme, and LSTM-based recovery scheme, respectively. The MI peak height and position clearly show improved performance for the LSTM-based scheme, achieving 74.7\% peak height recovery with no peak delay compared to 47.1\% peak height recovery and a 32.5~ns peak delay for the hardware-based scheme.
\label{fig:comparison}}
\end{figure*}

\subsection{Calculation of MI of Undisrupted Twin Beams}

To compute the MI, we use the same methodology described in Ref.~\cite{PRXQuantum.5.030351}. The intensity fluctuations of the twin beams are processed in Matlab using a band-pass filter set between 1.5~MHz and 3.5~MHz, where the quantum-correlated source exhibits optimal two-mode squeezing, as indicated by the shaded area in Fig.\ref{fig1}(b). Although the fluctuations in each beam's intensity appear random, the fluctuations in one beam provide information about those in the other. Consequently, the undisrupted probe and conjugate beams largely shared correlated information between them, surpassing the technical noise regime into the shot noise regime, i.e., beyond the SQL, as demonstrated by the similarities in the fluctuations of the undisrupted twin beams depicted in Figs.~\ref{fig:verification}(a) and~\ref{fig:verification}(b), respectively. As in Ref.~\cite{PRXQuantum.5.030351}, we use the following equation to capture these correlations~\cite{nielsen2010quantum}:  
\begin{equation}
    I(p;c) =  \sum\limits_{1}^{N_p} \sum\limits_{1}^{N_c} P(p,c) \log_2 \dfrac{P(p,c)}{P(p)P(c)},
    \label{eq1}
\end{equation}
where $P(p,c)$ is the joint probability of the probe ($p$) and conjugate ($c$) beams acquired from a 2-D histogram of the two beams' intensity fluctuations by \textit{individually binning the y-axes of their time sequences} in Figs.~\ref{fig:verification}(a) and (b). The two beams' marginal probabilities are denoted as $P(p)$ and $P(c)$, and $N_p$ and $N_c$ are the chosen number of bins for binning the two beams' intensity fluctuations on the $y-$axis.

We then calculate the MI as a function of time by shifting the two intensity fluctuations along the $x$-axis in increments of the oscilloscope's sampling resolution, i.e., 0.5~ns, and calculating Eq.~\eqref{eq1} at each time shift. It is important to note that Eq.~\eqref{eq1} calculates MI solely in the amplitude quadrature, unlike QMI, which necessitates knowledge of both quadratures of the two fields. Since we use $\text{log}_2$, the MI computed from Eq.~\eqref{eq1} is expressed in bits. 
%The number of bins can vary based on the dynamic range of the intensity fluctuations, which corresponds to the number of digitization levels occupied by the fluctuations. With a finite dynamic range, increasing the number of bins results in smaller bin sizes, allowing us to resolve finer, higher-frequency fluctuations. 
Considering that the majority of the probe and conjugate beams' intensity fluctuations  fall within 100 digitization levels, we choose $N_p=N_c=100$ for the data analysis to create the 2-D histograms.

Following this procedure, the MI of the original undisrupted twin beams as a function of time shift is depicted by the green curve in Fig.~\ref{fig:comparison}(c), corresponding to optical powers of 5.87~mW for the probe beam and 5.28~mW for the conjugate beam.

\subsection{Hardware Recovery of MI}

In the hardware-based scattering mitigation scheme outlined in Ref.~\cite{PRXQuantum.5.030351}, a scatterer is positioned in the path of the probe beam, followed by an IS, while the conjugate beam remains undisturbed. The probe beam's optical power before the scatterer is set at 5.87~mW, which drops to 740~$\mu$W after passing through the scatterer and being collected by the IS due to multiple non-unity reflections inside the IS. The conjugate beam's optical power is maintained at 5.28~mW. The time sequences of the disrupted probe and conjugate beams' intensity fluctuations are illustrated in Fig.~\ref{fig:comparison}(a) as the blue and green curves, respectively. As shown in Fig.~\ref{fig:comparison}(a), the dynamic range (i.e., digitization level) of probe intensity fluctuations is significantly reduced, corresponding to the significant drop in the probe beam's optical power (since variance is proportional to mean). In contrast, the dynamic range of intensity fluctuations for the conjugate beam remains unchanged as conjugate beam travels undisturbed. Additionally, since the scatterer inevitably introduces randomness to the probe intensity fluctuations, thereby obscuring the correlated fine structures with the conjugate beam that signify quantum correlations in the shot noise regime. This results in smoother intensity fluctuations of the disrupted probe beam after it travels through the scatterer and IS as can be seen in Fig.~\ref{fig:comparison}(a).

The recovered MI between the blue and green sequences in Fig.~\ref{fig:comparison}(a), calculated using Eq.~\eqref{eq1} for this hardware-based scheme, is depicted by the blue curve in Fig.~\ref{fig:comparison}(c). The peak height is reduced to 47.1~\% of its original value, with the peak position delayed by 32.5~ns relative to the green curve. \textcolor{black}{An intuitive explanation of the hardware-recovery result having a peak delay can be provided as follows. Because of the diffuser in the probe beam path, the probe photons are scattered nearly uniformly in all directions, covering a solid angle of $4\pi$ steradians. By placing an IS immediately after the diffuser, the forward-scattered probe photons from all directions are effectively recollected by the IS. These photons then undergo multiple reflections on the inner spherical surface of the IS before escaping through the exit port and eventually being focused on the photodiode. Due to these multiple reflections, a time delay in the MI peak relative to the unobstructed MI is observed.} %This time delay can be interpreted as the ``memory time" of the IS, suggesting that 47.1~\% of the undisrupted MI is stored in the IS for 32.5~ns.
We also examined a configuration with \textit {only the scatterer is present} in the probe beam path but \textit{without the IS}. In this scenario, The optical power of the disrupted probe beam is inadequate to produce significant intensity fluctuations, as these are dominated by the electronic noise of the detector, hence the MI is significantly diminished as represented by the black curve in Fig.~\ref{fig:comparison}(c).

\subsection{LSTM Recovery of MI and Two-Mode Squeezing}

Using the LSTM architecture described in Section~\ref{sec:lstm_arch}, the recovered probe beam time sequence is plotted along with the undisrupted conjugate beam time sequence as the red and green curves, respectively, in Fig.~\ref{fig:comparison}(b). Compared to the blue curve in Fig.~\ref{fig:comparison}(a) for the hardware-based scheme, the similarities between the LSTM-recovered probe sequence and the conjugate sequence in the presence of the scatterer are significantly improved. The corresponding MI between the red and green sequences in Fig.~\ref{fig:comparison}(b), calculated using Eq.~\eqref{eq1} is depicted in Fig.~\ref{fig:comparison}(c) by the red curve. The LSTM-based scheme achieves a MI peak height recovery of 74.7\% with no peak delay, compared to 47.1\% MI peak height recovery and a 32.5~ns peak delay in the hardware-based scheme.

\begin{figure}
\centering
\includegraphics[width=0.9\linewidth]{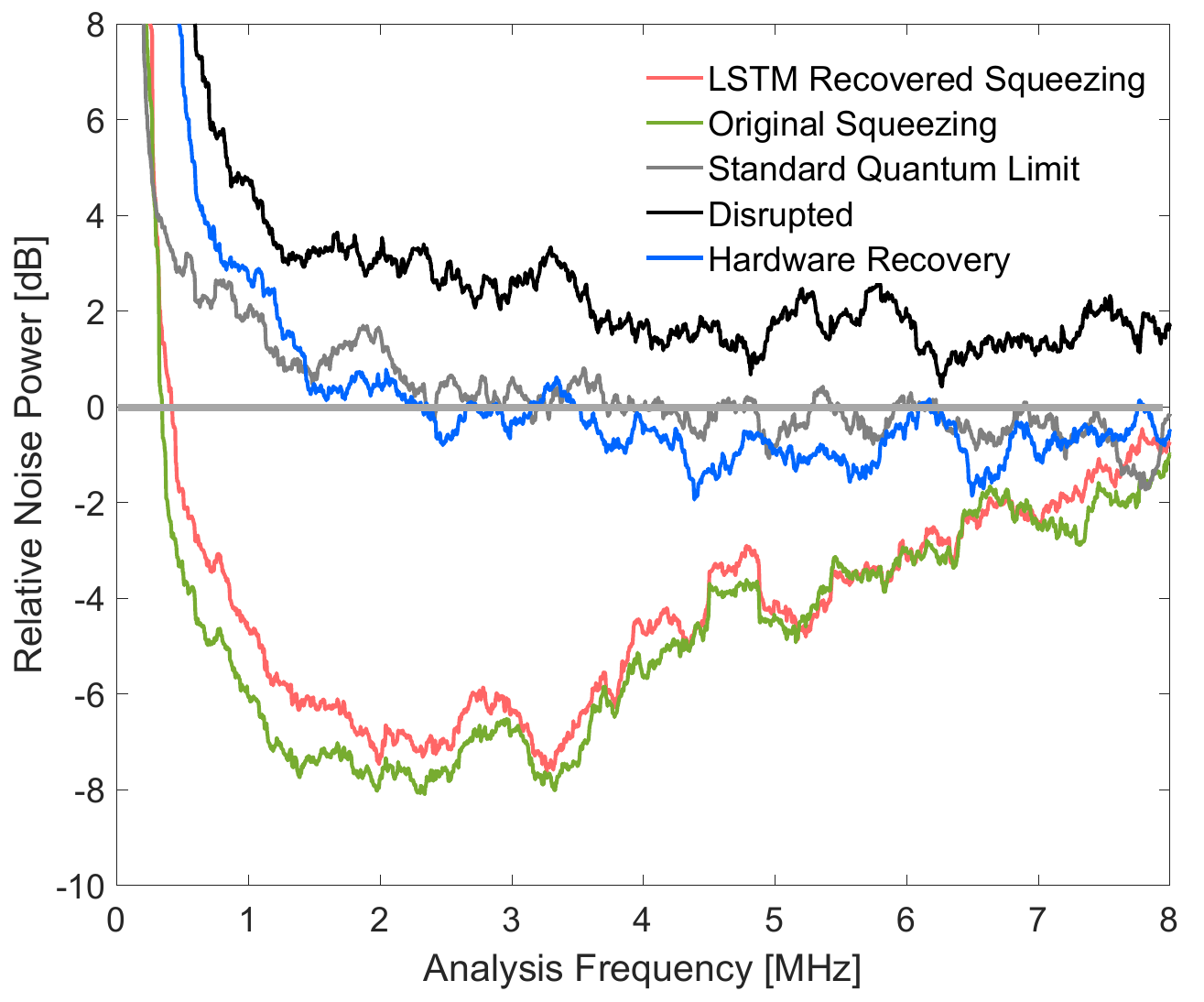}
\caption{Restoration of two-mode squeezing using the LSTM-based architecture. The noise power spectra for the standard quantum limit, original undisrupted squeezing, and LSTM-recovered squeezing are shown as the gray, green, and red curves, respectively. \textcolor{black}{The noise power spectra for the hardware recovered case and disrupted case are also shown as the blue and black curves respectively.} The LSTM architecture achieves an 87.7\% squeezing recovery within the analysis frequency from 1.5~MHz to 3.5~MHz.}
\label{fig:squeezing}
\end{figure}

Since squeezing is a direct indicator of quantum correlations, we also investigate the potential for squeezing recovery using the LSTM architecture. \textcolor{black}{We acknowledge that while intensity-difference squeezing provides strong evidence of quantum correlations, it may not capture all aspects of \textit{quantum coherence.}  The choice to measure squeezing in intensity fluctuations was driven by the specific goals of this study, which focused on mitigating the impact of scattering on quantum correlations. Intensity-difference squeezing offers a robust and experimentally accessible metric, particularly for two-mode quantum systems, although some quantum coherence information might not be fully reflected in intensity-difference squeezing alone. For instance, phase coherence between the modes, which requires additional measurements (e.g., field quadrature measurements), might reveal further insights into the quantum state.  In our study, we relied on well-established physical models and experimental setups (as detailed in Ref.~\cite{PRXQuantum.5.030351}) to ensure that the observed squeezing is directly tied to the underlying quantum correlations. Adding measurements that capture a broader range of quantum coherence properties could be an interesting avenue for future work. This might involve using covariance matrix reconstructions for direct measurements of quantum mutual information~\cite{clark2014quantum}.} 

We calculate the noise power spectrum between the LSTM-recovered probe sequence and the conjugate sequence in the presence of the scatterer, and plot the resulting curve as the red curve in Fig.~\ref{fig:squeezing}. The red curve, which is above the original undisrupted squeezing level (green curve) but below the standard quantum limit (shot noise level, oscillatory gray curve around 0 dB), indicates a successful recovery of quantum correlations. This suggests that our LSTM architecture effectively restores quantum correlations even after severe disruptive processes that would otherwise completely destroy them. \textcolor{black}{We also plot the relative noise power for the hardware recovered case and disrupted case in Fig.~\ref{fig:squeezing} as the blue and black curves, respectively. In the disrupted case, the scatterer nearly destroys the probe beam entirely, thus the intensity difference between the probe and conjugate beams is primarily determined by the conjugate beam, which alone exhibits super-Poissonian noise statistics. This results in a noise power exceeding the shot noise level, despite the optical power being only half that of the twin beams. In the hardware-recovered case, more than 80~\% photon loss causes the optical noise power of the recovered probe beam to be noticeably influenced by the probe detector's electronic noise floor, resulting in an intensity-difference noise power only slightly below the shot noise level.} To quantify the squeezing recovery, we calculate the average squeezing level between 1.5~MHz and 3.5~MHz, achieving an 87.7\% recovery compared to the green curve. The detailed experimental procedure for measuring the two-mode intensity-difference squeezing is provided in Ref.~\cite{li2021quantum}.

\section{Conclusion and Discussion}

We present an LSTM architecture designed to mitigate the negative impact of scattering on quantum correlations within a quantum system. In our approach, we utilize MI as the information metric for assessing quantum correlations. The system under consideration features a two-mode bright squeezed state, exhibiting nearly 8~dB of squeezing in the quantum-correlated twin beams' intensity difference, generated in a warm atomic vapor cell of $^{85}$Rb through the FWM process. By employing the LSTM-based scattering mitigation scheme, we demonstrate a 74.7~\% recovery of MI, a substantial improvement over the 47.4~\% hardware-based recovery reported in Ref.~\cite{PRXQuantum.5.030351} using an IS. We also demonstrate 87.7~\% recovery of the two-mode squeezing despite enormous photon loss that otherwise nullifies quantum correlations.

\textcolor{black}{\textit{Distinction from the hardware method.} The hardware method~\cite{PRXQuantum.5.030351} physically recovers quantum correlations by recollecting scattered photons, directly mitigating photon loss. In contrast, our ML-based approach reconstructs disrupted \textit {evolving} quantum correlations in a quantum system (rather than the full static state reconstruction in ML-aided QST) based on prior correlations and system characteristics. The distinction between the hardware and software methods indeed resemblances the difference between quantum error correction (which protects real quantum states) and quantum error mitigation (which focuses on correcting measurement outcomes). While it is true that our software-based approach may not directly address quantum communication tasks, which require transmission of physical quantum states with high fidelity, its utility should not be underestimated for other applications. Specifically, 
%the software-based method offers potential for certain pre-transmission tasks, such as noise mitigation and preparation of quantum resources, even if not suitable for direct quantum state transmission. Also, 
the software-based method eliminates the need to localize the scatterer, which is particularly advantageous in scenarios where environmental disturbances are unpredictable or cannot be directly addressed through hardware approach, such as in atmospheric quantum communication. Rather than being viewed as mutually exclusive, the hardware and software methods can \textit{complement} each other. For example, hardware approaches may handle primary recovery of lost photons, while software methods further enhance system resilience by mitigating residual decoherence effects.}

\textcolor{black}{\textit{Potential applications.} While our software approach does not physically recover lost photons, it serves as a powerful tool to infer and reconstruct quantum correlations in a quantum system in scenarios where physical recovery is either impractical or impossible. For example, in situations with severe scattering or inaccessible environments, our ML-based method provides a means to restore useful quantum correlations without hardware modifications. Specifically, the ML-based method can be employed in: (1) Scenarios with high loss: where physical recovery of photons is infeasible, such as in distributed quantum networks or highly scattering environments (e.g., atmospheric quantum communication); (2) Measurement error mitigation: by extrapolating missing information based on known system dynamics, the method can enhance the robustness of quantum systems for tasks like sensing or pre-transmission signal preparation; and (3) Cost-effective solutions: the software-based approach reduces the need for complex hardware modifications, making it an accessible alternative for resource-limited settings.} 

\textcolor{black}{It is worth noting that our LSTM-based ML approach can extend dynamical decoupling (DD), originally designed to mitigate decoherence and dissipation in open quantum systems~\cite{viola1999dynamical}, to quantum communication by predicting optimal pulse sequences. By analyzing real-time environmental noise and quantum correlation evolution, the LSTM model can dynamically adjust DD pulse timing and strength, potentially enhancing the performance of standard sequences like Carr-Purcell-Meiboom-Gill and Uhrig~\cite{ezzell2023dynamical}. This adaptive control can mitigate qubit dephasing~\cite{roy2012dynamical} and entanglement decay~\cite{barge2025polarization} in fiber-based quantum communication, providing software-driven noise suppression that improves reliability and optimizes resource efficiency without additional hardware modifications.} \\

%Scattering can induce decoherence and loss, which are the two dominate disruptive processes in quantum systems. Our LSTM-based scheme is particularly beneficial in situations where these disruptions are unavoidable, enabling the effective implementation of sensitive QISE protocols in practical applications. For instance, scattering would cause the deviations of quantum signals from their intended paths or reduce their fidelity in quantum communication systems, thus by mitigating scattering, one can preserve the coherence of quantum signals over long distances, which is essential for quantum key distribution~\cite{valivarthi2016quantum} and teleportation protocols~\cite{lu2022quantum}. In quantum computing platforms, where precise qubit manipulation is critical, scattering would induce errors by prematurely decohering qubits~\cite{krojanski2006reduced}, therefore reducing the effects of scattering is crucial for enhancing qubit fidelity and coherence times, ultimately improving the performance of quantum algorithms~\cite{schlosshauer2019quantum}. Moreover, scattering can interfere with ultra-sensitive measurement schemes that depend on quantum states in quantum sensing systems~\cite{matsuzaki2011magnetic}, such as in quantum metrology utilizing interferometric methods, mitigating scattering is vital for accurate signal detection, which would consequently enhance the signal-to-noise ratio of quantum sensors~\cite{taylor2016quantum}.\\

\section*{acknowledgments}

We thank the UTC Research Institute Center of Excellence in Applied Computational Science and Engineering (CEACSE-24) for supporting this work.

% This work supported by the U.S. Department of Energy, Office of Science, Office of Biological and Environmental Research under Award Number DE-SC-0023103 and DE-AC36-08GO28308. 

% \appendix

% \section{Appendixes}

% For binning of time traces, 


%\nocite{*}

\bibliography{MLQScattering}
\end{document}